\documentclass{svproc}
\usepackage{url}

\begin{document}
	\mainmatter              
	\title{Generalized uncertainty principle in bar detectors of gravitational waves}
	\titlerunning{GUP in resonant detectors of GWs}  
	%
	\author{Sukanta Bhattacharyya\inst{1} \and Sunandan Gangopadhyay\inst{2}
		\and Anirban Saha\inst{1}}
	\authorrunning{Sukanta Bhattacharyya et al.} 
	%
	\tocauthor{Sukanta Bhattacharyya,  Sunandan Gangopadhyay, Anirban Saha}
	\institute{ Department of Physics, West Bengal State University, Barasat, Kolkata 700126, India\\
		\email{sukanta706@gmail.com,} \email{sunandan.gangopadhyay@gmail.com, sunandan.gangopadhyay@bose.res.in}
		\and
	Department of Theoretical Sciences, S.N. Bose National Centre for Basic Sciences, JD Block, Sector III, Salt Lake, Kolkata 700106, India}

	\maketitle              
	
	\begin{abstract}
		At present the gravitational waves detectors achieve the sensitivity to detect the length variation ($\delta L$), $\mathcal{O} \approx 10^{-17}-10^{-21}$ meter. Recently a more stringent upperbound on the dimensionless parameter $\beta_0$, bearing the effect of generalized uncertainty principle has been given which corresponds to the intermediate length scale $l_{im}= \sqrt{\beta_0} l_{pl} \sim 10^{-23} m$. Hence it becomes quite obvious to search for the generalized uncertainty principle by observing the response of the vibrations of phonon modes in such resonant detectors in the near future. Therefore, we calculate the resonant frequencies and transition rates induced by the incoming gravitational waves on these detectors in the generalized uncertainty principle framework. This presentation is based on the work published in \cite{sb2}.
		
	\end{abstract}

\noindent The existence of an observer independent minimum length ($l_{pl} \approx 10^{-33} cm$), proposed by various quantum theories of gravity, demands a modification of the Heisenberg uncertainty principle (HUP) to the generalized uncertainty principle (GUP). Near the Planck energy scale, the Heisenberg uncertainty principle $\Delta x \sim \hbar/ \Delta p$ is expected to get modified by gravitational effects. The above observation along with various thought experiments leads to the following simplest form of the GUP \cite{kempf}-\cite{rm}  
\begin{eqnarray}
\Delta q_{i}\Delta p_{i}~\geq~\frac{\hbar}{2}\left[1+\beta(\Delta p^{2}+\langle
p\rangle^{2})+2\beta(\Delta p_{i}^{2}+<p_{i}>^{2})\right];~~~~i=1,2,3 \label{deltaxi}
\label{GUP}
\end{eqnarray}
where $p^2= \Sigma^3_{j=1} p_j p_j$ and $q_j$, $p_j$ are the position and its conjugate momenta. Naturally a lot of effort has been put to find an upper bound of the GUP parameter $\beta$ as it plays a crucial role for realizing the effects of GUP \cite{bawaj}. Again the testing of the GUP is extremely challenging and therefore initiates the proposal of a realistic experimental set up to test the GUP. On the other hand present day bar detectors \cite{GW-detection_status} are capable of detecting the fractional variations $\Delta L$ of the bar-length $L$ down to $\frac{\Delta L}{L} \sim 10^{-19}$, with $L\sim 1$ meter, which may be sensitive enough to allow us to probe the effects of quantum gravity. Motivated by the above discussion, in this presentation, we shall look at the quantum mechanical effects of the gravitational wave (GW) resonant bar detectors in the GUP framework. Physically these detectors can be described by a quantum mechanical forced GW-HO system. Hence we construct the quantum mechanical description of the GW-HO interaction in presence of the GUP. To carry this out we first write down the Hamiltonian of the GW-HO system as
\begin{equation}
{H} = \frac{1}{2m}\left({p}_{j} + m {\Gamma^j}_{0k} {q}^{k}\right)^2 + \frac{1}{2} m \varpi^{2} q_{j}^2  .
\label{e9}
\end{equation}
\noindent Following the standard prescription of quantum mechanics, we lift the phase-space variables  $\left( q^{j}, p_{j} \right)$ to operators 
$\left( {\hat q}^{j}, {\hat p}_{j} \right)$ in the GUP framework. 
These can be defined upto first order in $\beta$ as 
\begin{eqnarray}
\hat{q}_{i}={q}_{0i}~,~~~~~\hat{p}_{i}={p}_{0i}(1+\beta {p}_{0}^{2}) ~.
\label{GUPR}
\end{eqnarray}
\noindent Here we shall consider the resonant bar detectors as a $1$-D \footnote{A typical bar is a cylinder of length $L=3$ m and radius $R = 30$ cm, therefore one can treat its vibrations as one-dimensional.} system \cite{Magg}. Now in terms of the standard definition of the raising and lowering operators ($a, a^\dagger$), the Hamiltonian in eq.(\ref{e9}) up to first order in $\beta$ can be recast as $H=H_{0}+H_{1}+H_{2}$, where
\begin{eqnarray}
H_0 &=& \hbar \omega\left(a^\dagger a +\frac{1}{2}\right)~, \nonumber
\end{eqnarray}
\begin{eqnarray}
H_{1} &=& \frac{\beta}{m}\left(\frac{\hbar m \omega}{2}\right)^2\left[aaaa-aaaa^\dagger-aaa^\dagger a+aaa^\dagger a^\dagger-aa^\dagger aa+aa^\dagger a a^\dagger+aa^\dagger a^\dagger a \right. \nonumber\\ &&  \left. -a a^\dagger a^\dagger a^\dagger  
-a^\dagger aaa+a^\dagger aa a^\dagger +a^\dagger a a^\dagger a -a^\dagger a a^\dagger a^\dagger +a^\dagger a^\dagger a a-a^\dagger a^\dagger a a^\dagger \right. \nonumber\\ && \left. -a^\dagger a^\dagger a^\dagger a+a^\dagger a^\dagger a^\dagger a^\dagger\right]~,\nonumber
\end{eqnarray}
\begin{eqnarray} 
H_{2} &=& i \hbar \dot h_{11}\left[-\frac{1}{2}\left(aa-a^\dagger a^\dagger\right)+\frac{\beta \hbar m \omega}{4}\left(aaaa-aa a^\dagger a-aa^\dagger aa+a a^\dagger a^\dagger a
-a^\dagger aa a^\dagger \right. \right.
\nonumber\\ && \left. \left. +a^\dagger a a^\dagger a^\dagger +a^\dagger a^\dagger a a^\dagger  -a^\dagger a^\dagger a^\dagger a^\dagger\right)\right] ~.
\label{14}
\end{eqnarray}
\noindent Here $H_0$ stands for the Hamiltonian of ordinary HO while $H_{1}$ and $H_{2}$ are the time independent and time dependent part of the Hamiltonian respectively. Now we proceed to calculate the perturbed eigenstates and energies due to time independent Hamiltonian $H_1$. Using time independent perturbation theory, the perturbed eigenstates and their corresponding energies read
\begin{eqnarray}
| n\rangle^\beta &=& |n \rangle + \Delta \left[\frac{(2n+3)\sqrt{(n+1)(n+2)}}{4} |n+2\rangle - \frac{(2n-1)\sqrt{n(n-1)}}{4} |n-2 \rangle  \right. \nonumber\\ && \left.  +\frac{\sqrt{n(n-1)(n-2)(n-3)}}{16}|n-4\rangle-\frac{\sqrt{(n+1)(n+2)(n+3)(n+4)}}{16}|n+4\rangle \right]\nonumber
\label{9}
\end{eqnarray}
and
\begin{eqnarray}
E_n^{(\beta)} =\left(n+\frac{1}{2} \right) \hbar \omega \left[1+\frac{3(2n^2+2n+1)}{2(2n+1)}\Delta \right]~.
\label{en}
\end{eqnarray}
Here $\Delta= \beta \hbar m \omega$ is the dimensionless parameter involving the GUP parameter. Now the time dependent part of the Hamiltonian, that is, $H_2$ gives the probability amplitude of transition to be
\begin{eqnarray}
C_{0^\beta \rightarrow 2^\beta } =A \int_{-\infty}^{t\rightarrow +\infty} dt'~~ \dot h_{11}~~e^{i(2+9 \Delta)\omega t'} \nonumber\\
C_{0^\beta \rightarrow 4^\beta} =B \int_{-\infty}^{t\rightarrow +\infty} dt'~~ \dot h_{11}~~e^{i(4+30 \Delta)\omega t'}
\label{10}
\end{eqnarray}
where $A= \left(\frac{1}{\sqrt{2}}+\frac{9}{4 \sqrt{2}} \Delta\right)$ and{\bf{ $B= -3 \sqrt{6}\Delta $ }}are dimensionless constants. Eq.(\ref{10}) is one of the main results \cite{sb2}. From the transition amplitudes the presence of the GUP can be checked by measuring the corresponding transition probabilities $P_{i\rightarrow f}=  |C_{i\rightarrow f}|^{2}$. We now look at some GW templates. In this presentation we explore the effects of GUP in GWs detection technique using two types of GW templates generated from different astronomical events. First we consider the periodic GW with linear polarization. This has the form
\begin{equation}
h_{jk} \left(t\right) = 2f_{0} \cos{\Omega t} \left(\varepsilon_{\times}\sigma^1_{jk} + \varepsilon_{+}\sigma^3_{jk}\right)
\label{lin_pol}
\end{equation}
where the amplitude varies sinusoidally with a single frequency $\Omega$. The transition rates in this case are
\begin{eqnarray}
\lim\limits_{T \rightarrow \infty} \frac{1}{T}P_{0^\beta\rightarrow 2^\beta} &= &  \left(2 \pi f_0 \Omega A \epsilon_+\right)^2 \times \left[\delta\left( \omega \left(2 +9  \Delta \right)- \Omega \right)  \right]
\label{tm} \\
\lim\limits_{T \rightarrow \infty} \frac{1}{T}P_{0^\beta\rightarrow 4^\beta} &=&  \left(2 \pi f_0 \Omega B \epsilon_+\right)^2 \times \left[ \delta \left(  \omega \left(4 +30 \Delta \right) -\Omega \right)  \right]~.
\label{trlpgw}
\end{eqnarray}
\noindent With the above results in place, we now make an estimate of the GUP parameter $\beta_0$. The inequality $9\Delta\omega < 2\omega$ gives $ \beta_0 < 1.4 \times 10^{28}$~, $\beta = \beta_{0}/(M_{pl} c)^2$.
This upper bound on the GUP parameter is stronger than the one obtained in \cite{fm}.
Interestingly, we find that the correction to the resonant frequency $2\omega$ due to the GUP is $9\Delta\omega/(2\pi) \approx 1.3~kHz$. Hence, this simple calculation shows that the GUP modes can ring up in order to be detected by the resonant bar detectors. 

\noindent We shall next take approximated models of the astrophysical phenomenon like bursts which emits aperiodic GWs waveform with modulated Gaussian function as the second GW template. This has the form
\begin{eqnarray}
h_{jk} \left(t\right) = 2f_{0} g \left( t \right) \left(\varepsilon_{\times}\sigma^1_{jk} + \varepsilon_{+}\sigma^3_{jk}\right);~~~~ g \left(t\right) = e^{- t^{2}/ \tau_{g}^{2}}  \,  \sin \Omega_{0}t~.
\label{lin_pol_burst}
\end{eqnarray}
\noindent The transition probabilities for this case are
\begin{eqnarray}
P_{0^\beta\rightarrow 2^\beta} \approx e^{- \left(2 \omega +9 \omega \Delta-\Omega_{0}\right)^{2}\tau_{g}^{2}/2} \times  \left\{f_{0} \epsilon_+ A \sqrt{\pi}\tau_{g}\left(2 \omega+9\omega \Delta\right) \right\}^2
\nonumber\\
P_{0^\beta\rightarrow 4^\beta} \approx e^{- \left(4 \omega +30 \omega \Delta-\Omega_{0}\right)^{2}\tau_{g}^{2}/2}  \times  \left\{f_{0} \epsilon_+ B \sqrt{\pi}\tau_{g}\left(4 \omega+30\omega \Delta\right) \right\}^2~.
\label{gtdf22}
\end{eqnarray}
\noindent Now we proceed to conclude our presentation by summarizing the results.
We make the following observations from the exact forms of the transition rates. The resonant frequencies $\Omega=\omega(2+9\Delta)$ and $\Omega=\omega(4+30\Delta)$ get modified by the GUP parameter $\beta$.  This observation is quite similar with that obtained in the noncommutative framework \cite{sg4},\cite{sg5}. In the presence of the GUP, we find that there are more than one transitions possible from the ground state to the excited states with different amplitudes. We get both the linear as well as the quadratic terms in the dimensionless  GUP parameter $\Delta$ in the expressions of the transition amplitudes. The linear dependence in $\Delta$ is easier to detect. We have also done this analysis for the circularly polarised GWs which show that they are also good candidates to probe the presence of the GUP in the resonant detectors \cite{sb2}. In this presentation, we have shown an upper bound estimation of the GUP parameter $\beta_0<10^{28}$. This is a much stronger bound than that obtained in \cite{fm} which is $\beta_0<10^{33}$. The observations made here reveal that resonant detectors may allow in the near future to detect the existence of an underlying generalized uncertainty principle. Moreover, in the recent literature \cite{sb}, a connection between the generalized uncertainty principle and the spatial noncommutative structure of space has been shown. Our analysis also indicates a similarity between the findings in these two frameworks.



\end{document}